\documentclass[10pt,aps,prd,showpacs,amsmath,amssymb,nofootinbib,twocolumn,longbibliography,superscriptaddress]{revtex4-1}

\usepackage[english]{babel}				
\usepackage[mathscr]{eucal} 			
\usepackage{mathtools}					
\usepackage[ddmmyyyy]{datetime}    		
\usepackage{enumitem}					
\usepackage[dvipsnames]{xcolor}			

\usepackage{esint}

\usepackage[colorlinks=true,
            citecolor=ForestGreen,
            linkcolor=red,
            filecolor=cyan,
            urlcolor=magenta,
            backref=false]{hyperref}

\AtBeginDocument{%
    \newwrite\bibnotes
    \def\bibnotesext{Notes.bib}
    \immediate\openout\bibnotes=\jobname\bibnotesext
    \immediate\write\bibnotes{@CONTROL{REVTEX41Control}}
    \immediate\write\bibnotes{@CONTROL{%
    apsrev41Control,author="08",editor="1",pages="1",title="0",year="1"}}
     \if@filesw
     \immediate\write\@auxout{\string\citation{apsrev41Control}}%
    \fi
}%

\newcommand\bs[1]{\boldsymbol{#1}}
\newcommand{\ts}[1]{{\boldsymbol{#1}}}

\newcommand\dd{\mathrm{d}}

\newcommand\feq{\mathrel{\phantom{=}}}

\DeclareMathOperator{\sign}{sign}

\DeclareMathOperator\Ei{Ei}

\DeclareMathOperator\artanh{artanh}

\DeclareMathOperator\Det{Det}

\begin{document}


\title{Retarded field of a uniformly accelerated source in non-local scalar field theory}

\author{Ivan Kol\'a\v{r}}
\email{i.kolar@rug.nl}
\affiliation{Van Swinderen Institute, University of Groningen, 9747 AG, Groningen, The Netherlands}

\author{Jens Boos}
\email{jboos@wm.edu}
\affiliation{High Energy Theory Group, Department of Physics, William \& Mary, Williamsburg, VA 23187-8795, United States}

\date{\today}

\begin{abstract}
We study the retarded field sourced by a uniformly accelerated particle in a non-local scalar field theory. While the presence of non-locality regularizes the field at the location of the source, we also show that Lorentz-invariant non-local field theories are particularly sensitive to the somewhat unphysical assumption of uniform acceleration, leading to logarithmic divergences on the acceleration horizon. Analytic properties of the non-local retarded Green function indicate that the divergences can be removed by placing appropriate sources on the acceleration horizon in the asymptotic past.
\end{abstract}

\maketitle

\section{Introduction}

Locality is deeply woven into our notion of physics: from classical mechanics to general relativity and quantum field theory, locality has been an undergirding principle across disciplines. However, there are notable exceptions from that rule. Quantum entanglement is a non-local phenomenon, effective actions in quantum field theory typically contain non-local factors, and it has proven difficult if not outright impossible to define local observables in quantum gravity \cite{Kiefer:2014}. Therein, the role of non-locality may also play a major role in possible resolutions of the black hole information loss problem \cite{Giddings:2012gc}.

The recent years have seen a flurry of activity with a particular focus on the class of ghost-free infinite-derivative theories \cite{Tomboulis:1997gg,Modesto:2011kw,Biswas:2011ar}. These theories propose a fundamental non-locality by means of non-local form factors $f(\Box)$, and have been remarkably successful in alleviating curvature singularities \cite{Frolov:2015usa,Frolov:2015bia,Frolov:2015bta,Edholm:2016hbt,Boos:2018bxf,Buoninfante:2018rlq,Buoninfante:2018stt,Boos:2020ccj,Kolar:2020bpo} in the context of weak-field gravity. Some exact non-singular solutions of infinite-derivative gravity theories have been constructed in the context of gravitational waves \cite{Kilicarslan:2019njc,Dengiz:2020xbu} and cosmology \cite{Biswas:2005qr,Biswas:2010zk}. Implications of such non-local modifications have also been investigated in quantum theory \cite{Boos:2018kir,Buoninfante:2019teo,Boos:2020hxu}, quantum field theory \cite{Shapiro:2015uxa,Carone:2016eyp,Briscese:2018oyx,Buoninfante:2018mre,Boos:2019fbu}, quantum field theory in curved spacetime \cite{Boos:2019vcz}, Hamiltonian mechanics \cite{Calcagni:2007ef,Kolar:2020ezu}, and other aspects of gravitational theory \cite{Calcagni:2010ab,Modesto:2017sdr,Kolar:2020max}. Non-local Green functions have proven a particularly useful tool in such studies \cite{Boos:2020qgg}, even though most scenarios considered in the literature so far are either time-independent or space-independent, implying that the full spacetime notion of non-local Green functions is not yet very well understood.

This paper aims towards closing that gap by studying the retarded non-local scalar field of a uniformly accelerated source in flat spacetime. The study of the retarded field for uniformly and arbitrarily accelerated point particles has a long history, but, to the best of our knowledge, has so far been focused on local field theories.

In 1909, Born studied the field of two charges undergoing uniform acceleration in opposite directions \cite{Born:1909}. The following decades saw substantial activity in this field, and while much progress was made in analyzing the radiation content of such a field configuration---see e.g. the introduction in Fulton and Rohrlich \cite{Fulton:1960} for a brief historical overview---Bondi and Gold \cite{Bondi:1955zz} emphasized that the behavior of the field on the acceleration horizons was singular. Boulware \cite{Boulware:1979qj} and Das \cite{Das:1980} considered physically meaningful limiting procedures towards the unphysical assumption of uniform acceleration, and Bondi \cite{Bondi:1981} used their approach to re-derive the original Bondi--Gold solution. While Ginzburg has deemed the problem of the radiation of uniformly accelerated charges solved \cite{Ginzburg:1970,Ginzburg:1979,Ginzburg:1989}, the field is still active, focusing on the influence of gravitation \cite{Zelnikov:1982}, studying scalar theory \cite{Ren:1993bs}, or extending the studies to de Sitter spacetime \cite{Bicak:2002yk,Bicak:2005yt}.

These considerations have provided much insight on the causal structure of fields propagating in Minkowski spacetime, the spacetime properties of retarded Green functions, and have brought to light some unphysical consequences of assuming uniform acceleration. This paper presents a first step towards extending many of these considerations from local field theory to a class of \emph{non-local} field theories.

In order to focus our discussion somewhat we shall consider a simple toy model of a scalar field theory in four-dimensional Minkowski spacetime with the metric
\begin{align}
\dd s^2 = g{}_{\mu\nu} \dd X{}^\mu \dd X{}^\nu = -\dd t^2 + \dd x^2 + \dd y^2 + \dd z^2 \, ,
\end{align}
expressed in Cartesian coordinates $X{}^\mu = (t,\ts{x})$ where we denoted $\ts{x}=(x,y,z)$ for simplicity. The scalar field equation takes the simple form
\begin{align}
\mathcal{D} \phi = j \, ,
\end{align}
where $j$ is an external source, and $\mathcal{D}$ is a differential operator.\footnote{We use the letter $j$ to denote the external source term, but recall that a scalar field theory couples to a density and \emph{not} to a conserved current.} The \emph{local theory} is specified by the choice $\mathcal{D} = \Box$, where $\Box$ is the d'Alembert operator, and one recovers the massless Klein--Gordon equation. Suppose now that the external source has the following form:
\begin{align}
\begin{split}
j(X) &=2\mu\alpha\delta{}^{(2)}(-t^2+z^2-\alpha^2)\delta(x)\delta(y) \\
&\hspace{11pt} \times \theta(z+t)\theta(z-t) \, ,
\end{split}
\end{align}
which describes a uniformly accelerated particle of mass $\mu>0$ and acceleration parameter $\alpha$ such that the constant acceleration of the particle is $\mu/\alpha$, and the particle is located on the positive part of the $z$-axis. The retarded field created by such a source may be calculated via the retarded Green function
\begin{align}
\begin{split}
G{}^\text{R}(X',X) &= \frac{1}{2\pi} \delta{}^{(2)}\left[(X'-X)^2\right] \theta(t'-t) \, , \\
(X'-X)^2 &= -(t'-t)^2 + (\ts{x}'-\ts{x})^2 \, ,
\end{split}
\end{align}
such that the retarded solution for $\phi$ takes the well known form \cite{Zelnikov:1982,Ren:1993bs,Bicak:2002yk,Bicak:2005yt}
\begin{align}
\begin{split}
\phi(X) &= \int \dd^4 X' G^\text{R}(X,X') j(X') \\
&= -\frac{\mu\alpha}{2\pi}\frac{\theta(z+t)}{\sqrt{(X^2+\alpha^2)^2- 4\alpha^2 (z^2-t^2)}} \, .
\end{split}
\end{align}
This retarded field of a uniformly accelerated source has several remarkable properties.

First, this expression diverges when $-t^2+z^2=\alpha^2$ and $x=y=0$, that is, at the location of the uniformly accelerated source. Second, this expression is non-zero only in the future and right Rindler wedges, while being finite on all horizons. And third, across the past acceleration horizon located at $u\equiv z+t=0$, the retarded field exhibits a discontinuity:
\begin{align}
\Delta\phi^{u=0} \equiv \phi(u=0^+)-\phi(u=0^-) = -\frac{\mu}{2\pi\alpha} \, .
\end{align}
These three properties are intimately connected to the properties of the retarded Green function of the local scalar theory.

In the remainder of this paper it is our objective to understand how the presence of non-locality affects the properties of the retarded field of a uniformly accelerated particle. Our model of non-locality utilizes the following differential operator,
\begin{align}
\mathcal{D} = \exp\Big[(-\ell^2\Box)^N\Big]\Box \, , \quad N = 1, 2, \dots \, , \quad \ell > 0 \, .
\end{align}
This expression is to be understood via a formal expansion. $N$ is an integer, and $\ell>0$ is the \emph{scale of non-locality}, and this class of non-local theories is also referred to as $\mathrm{GF_N}$. Here, ``GF'' stands for ``ghost-free'' since the inverse of the non-local differential operator in Fourier space has no additional poles and is thereby devoid of spurious ghost-like particles typically encountered in higher-derivative theories. In the local limit $\ell\rightarrow 0$ one recovers the local theory. It has been demonstrated that $\mathrm{GF_N}$ theories manifestly regularize the field of stationary sources, but in the time-dependent case only even values for $N$ are permissible, since odd $N$ lead to time-dependent instabilities and divergences in the classical theory \cite{Frolov:2016xhq,Boos:2019fbu}.

Moreover, in a true spacetime sense it is impossible to define ``small'' Lorentz-invariant spacetime volumes by relations of the form $-(t'-t)^2+(\ts{x}'-\ts{x})^2 < \ell^2$ since they are always hyperbolic in nature. While in many purely spatial problems the question of time-dependence can be neglected and non-locality truly acts on a small scale, in the present paper this interpretation is not possible. For this reason we will place particular focus and emphasis on non-local effects close to the light cone.

This paper is organized as follows. In Sec.~\ref{sc:Mink} we will briefly introduce some useful coordinate systems and the notion of Fourier transforms in those curvilinear coordinate systems. In Sec.~\ref{sc:nlsol} we will derive an integral expression for the retarded field of a uniformly accelerated source in the non-local theory and discuss its properties in detail. And last, in Sec.~\ref{sc:con}, we will summarize our findings and outline possible future research directions.


\section{Minkowski spacetime}\label{sc:Mink}
In what follows it will be useful to work in Rindler coordinates, so let us briefly fix our notation to encompass different coordinate choices both in real space and Fourier space.

\subsection{Various coordinates}
\begin{figure}
    \centering
    \includegraphics[width=0.45\textwidth]{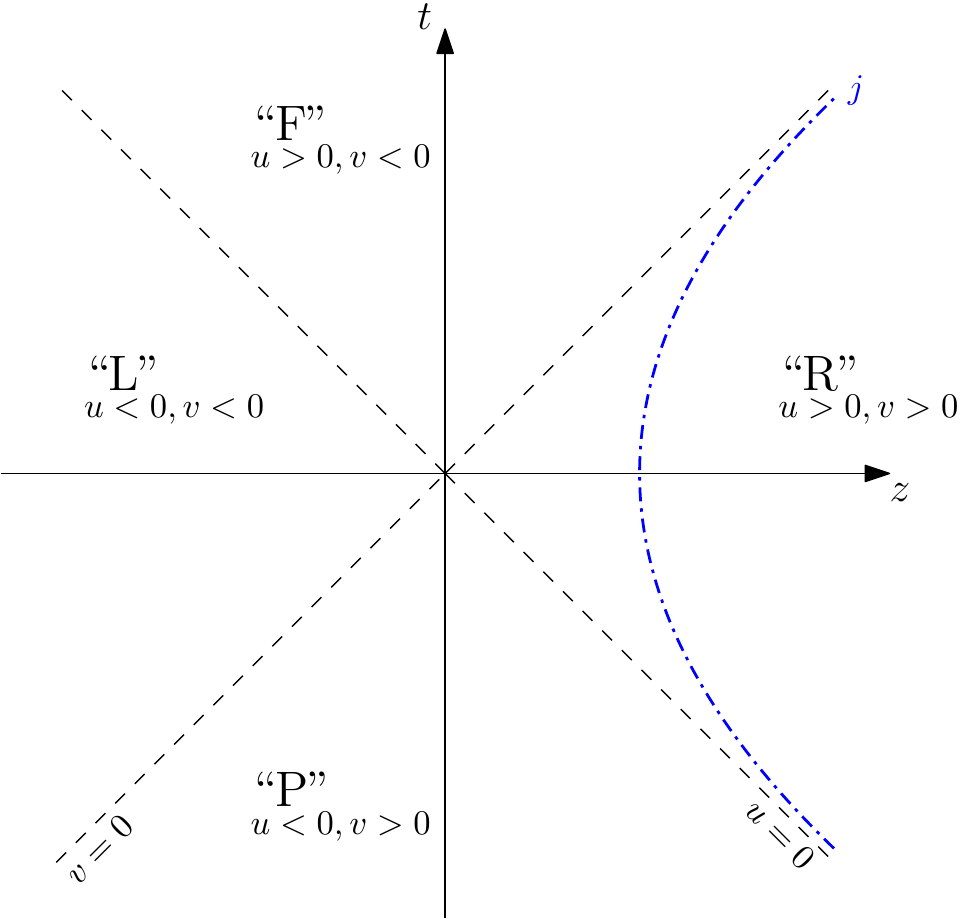}
    \caption{The split of Minkowski spacetime into the four regions ``L,'' ``R,'' ``F,'' and ``P,'' here displayed for $x=y=0$. The dash-dotted line represents the spacetime location of the uniformly accelerated particle, and the dashed line $v=0$ ($u=0$) represent the future (past) acceleration horizon.}
    \label{fig:wedges}
\end{figure}

In this paper we exclusively consider flat Minkowski spacetime, but it is convenient to introduce several coordinates. We start with the standard \textit{Cartesian coordinates} $\{t, x, y, z\}$, where the flat metric takes the form
\begin{align}
\dd s^2 = -\dd t^2 + \dd x^2 + \dd y^2 + \dd z^2 \, .
\end{align}
It is useful to transform to \textit{null coordinates} $\{u, v\}$ via\footnote{Note that in the literature one also finds the alternative definitions $\check{u} = t-z$ and $\check{v} = t+z$. We choose the present convention such that $u>0$ and $v>0$ in the right Rindler wedge, which reduces the amount of signs encountered in the following calculations significantly.}
\begin{align}
    u =z+t\,, \quad v =z-t \, .
\end{align}
Finally, let us define the the \textit{real Rindler coordinates} $\{\tau, \zeta, x, y\}$ that are adapted to the boost Killing vector $z\partial_t-t\partial_z$ such that
\begin{equation}
\begin{aligned}
    \tau &=\tfrac12\log|u/v|=\artanh\left[(t/z)^{\sigma_u\sigma_v}\right]\;,
    \\
    \zeta &=\sqrt{|uv|}=\sqrt{|-t^2+z^2|}\;,
\end{aligned}
\end{equation}
where $\sigma_u=\sign(u)$ and $\sigma_v=\sign(v)$. The inverse transformations are given by
\begin{alignat}{3}
    u &=\sigma_u \zeta e^\tau\;, &
    v &=\sigma_v \zeta e^{-\tau}\;,
    \\
    t&=\tfrac{\zeta}{2}(\sigma_u e^\tau-\sigma_v e^{-\tau})\;, & \quad
    z&=\tfrac{\zeta}{2}(\sigma_u e^\tau+\sigma_v e^{-\tau})\;.
\end{alignat}
Introducing the subscript ${\textrm{W}\in\{\textrm{R},\textrm{L},\textrm{F},\textrm{P}\}}$ we may label individual regions of Minkowski spacetime as $M_\mathrm{W}$; see Fig.~\ref{fig:wedges}. The metric  in Rindler coordinates is
\begin{equation}
\dd s^2 = \sigma_u\sigma_v\big({-}\zeta^2\dd\tau^2+\dd \zeta^2\big) + \dd\rho^2 + \rho^2\dd\varphi^2 \, ,
\end{equation}
where we also introduced the polar cylindrical version given by the standard relations $(x,y)=(\rho\cos{\varphi},\rho\sin{\varphi})$. Denoting the four-dimensional spacetime volume element by ${\mathfrak{g}^{1/2}=\sqrt{|\Det{\bs{g}}|}}$, we can write
\begin{equation}
    \mathfrak{g}^{1/2}=\dd t\dd x\dd y\dd z=\tfrac12\dd u\dd v \dd x\dd y = \zeta \rho \, \dd \tau \dd \zeta \dd\rho \dd\varphi\;.
\end{equation}
The norm of a position vector $\ts{X}$ in these coordinates reads
\begin{align}
\begin{split}
\ts{X}^2 &\equiv \ts{X}\cdot\ts{X}\equiv g_{\mu\nu} X^\mu X^\nu = -t^2+x^2+y^2+z^2 \\ 
&=uv+x^2+y^2=\sigma_u\sigma_v\zeta^2+\rho^2\;,
\end{split}
\end{align}
where the dot denotes the scalar product. The difference of two such vectors $\ts{X}$ and $\tilde{\ts{X}}$ has the norm
\begin{align}
(\ts{X}-\tilde{\ts{X}})^2 &=-(t-\tilde{t})^2+(x-\tilde{x})^2+(y-\tilde{y})^2+(z-\tilde{z})^2 \nonumber \\
    &=e^{-\tau-\tilde{\tau}} \big(\sigma_u e^\tau \zeta-\sigma_{\tilde{u}} e^{\tilde{\tau}} \tilde{\zeta}\big) \big(\sigma_v e^{\tilde{\tau}} \zeta-\sigma_{\tilde{v}} e^\tau \tilde{\zeta}\big) \nonumber \\
    &\hspace{11pt}+\rho^2 + \tilde{\rho}^2 - 2\rho\tilde{\rho}\cos(\varphi-\tilde{\varphi}) \, .
\end{align}

\subsection{Fourier transform}
Due to the translational invariance of Minkowski spacetime $M$ it is convenient to employ Fourier transform methods. Because Minkowski spacetime is an affine space, after fixing an arbitrary origin one may freely convert coordinate positions into vectors with respect to that origin. We denote the Fourier transform of a function $f(X)$ as $f_{\bar{X}}$, and in four spacetime dimensions their interrelations are given by the formulas
\begin{align}
f_{\bar{X}}&=\frac{1}{4\pi^2}\int\limits_{M}\!\mathfrak{g}^{1/2}(X) \, e^{+i\bar{\ts{X}}\cdot \ts{X}} \, f(X) \, , \\
f(X) &= \frac{1}{4\pi^2}\int\limits_{M}\!\mathfrak{g}^{1/2}(\bar{X})  e^{-i\bar{\ts{X}}\cdot \ts{X}} \, f_{\bar{X}} \, .
\end{align}
In our terminology, the \textit{coordinate space} vector $\ts{X}$ as well as \textit{momentum space} vector $\bar{\ts{X}}$ live in the same vector space. This definition is useful because now several coordinate systems can be used both for the Fourier transform and its inverse. The contraction between the momentum space and coordinate space vectors is given by
\begin{align}
\bar{\ts{X}}\cdot \ts{X} &=-\bar{t}t+\bar{x}x+\bar{y}y+\bar{z}z \\
&=\tfrac{\bar{\zeta}\zeta}{2}(\sigma_{\bar{u}}\sigma_v e^{\bar{\tau}-\tau}+\sigma_{\bar{v}}\sigma_u e^{-\bar{\tau}+\tau})+\bar{\rho}\rho\cos{(\bar{\varphi}{-}\varphi)} \, . \nonumber
\end{align}
Let us emphasize here that the above notation presents a departure from the common notation where one would write $X{}^{\mu} = (t,\ts{x})$ and $\bar{X}^\mu = (\omega,\ts{k})$. Hence, in what follows, a barred quantity is the conjugate Fourier momentum to the unbarred real-space variable. For Euclidean coordinates this procedure is somewhat odd, but its notational advantage becomes apparent when performing Fourier transforms in curvilinear coordinates---as we shall see below---since in that case one does not need to invent new coordinate symbols for Fourier space. For similar methods in Lorentz-invariant Fourier transforms we refer to the insightful paper by DeWitt-Morette \textit{et al.} \cite{Wurm:2003}.

\section{Non-local solution for an accelerated particle}\label{sc:nlsol}

\subsection{Non-local theory}

With the brief reminder on Lorentz-invariant Fourier transforms out of the way, let us discuss our non-local toy model. In what follows we shall consider a scalar field theory described by the equation of motion
\begin{equation}\label{eq:eomnonlocal}
a(\Box)\Box \phi = j \, .
\end{equation}
Here, $a(\Box)$ is an analytic operator
\begin{equation}
    a(\Box)=\sum_{k=0}^\infty a_k \Box^k \, ,
\end{equation}
and $j$ is an external source. In this paper, for simplicity, we focus on so-called $\mathrm{GF_N}$ theories defined by
\begin{equation}\label{eq:GFN}
    a(\Box)=\exp\Big[(-\ell^2\Box)^N\Big]\;,
\end{equation}
which reduces to the local case ${a(\Box)=1}$ in the \textit{local limit} ${\ell\to0}$. $a(\Box)$ is called \emph{form factor}, and it satisfies two important properties: it is non-vanishing when acting on functions, and it satisfies $a(0) = 1$.

\subsection{Retarded solution}
Consider a particle with mass ${\mu}$ that uniformly accelerates in the direction of the positive $z$-axis with the constant acceleration ${\mu/\alpha}$. The corresponding source is localized in the region $M_{\textrm{R}}$ and can be parametrized as
\begin{equation}\label{eq:source}
\begin{aligned}
    j(X) &=2\mu\alpha\delta{}^{(2)}(-t^2+z^2-\alpha^2)\delta(x)\delta(y)\theta(u)\theta(v)
    \\
    &=\mu\delta(\zeta-\alpha)\delta(x)\delta(y)\theta(u)\theta(v)\;.
\end{aligned}
\end{equation}
In order to find the response of the non-local theory to this source, we employ the Green function method such that the retarded solution is given by the integral
\begin{equation}\label{eq:retsolcongr}
\phi(X) = \frac{1}{4\pi^2}\int\limits_{M}\!\mathfrak{g}^{1/2}(\bar{X}) \, e^{-i\bar{\bs{X}}\cdot\bs{X}} ~ \mathcal{G}^\text{R}_{\bar{X}} \, j_{\bar{X}} \, .
\end{equation}
Real-space expressions for $\mathcal{G}^\text{R}$ are known and can be given in terms of Meijer-G functions, and we derive an explicit expression in Appendix~\ref{app:deltaG}, where we also prove that they satisfy DeWitt's asymptotic causality criterion \cite{DeWitt:1965}. However, their form is rather complicated and hence impractical for calculational purposes. As we will demonstrate now it is much simpler to perform the calculations in momentum space. 

A momentum space Green function for the differential operator $\Box a(\Box)$ is given by the expression
\begin{align}
\mathcal{G}_{\bar{X}} = \frac{1}{-\bar{\ts{X}}^2 a(-\bar{\ts{X}}^2)} \, ,
\end{align}
and we may rewrite it as a sum of two terms,
\begin{align}
\begin{split}
\label{eq:deltaG-causality}
\mathcal{G}_{\bar{X}} &= G_{\bar{X}} + \Delta\mathcal{G}_{\bar{X}} \, , \\
G_{\bar{X}} &= \frac{1}{-\bar{\ts{X}}^2} \, , \quad \Delta\mathcal{G}_{\bar{X}} = \frac{a^{-1}(-\bar{\ts{X}}^2)-1}{-\bar{\ts{X}}^2} \, .
\end{split}
\end{align}
Here, $G_{\bar{X}}$ denotes the Green function of the $\Box$-operator. This quantity is a Green function for the local theory and does not depend on the presence of non-locality. It has two poles in complex Fourier space, and needs to be regulated, typically via a suitable $i\epsilon$-prescription. As is well known, the choice of $i\epsilon$-regularization gives rise to distinct causal properties.
The quantity $\Delta\mathcal{G}_{\bar{X}}$, on the other hand, encapsulates the non-local modification of the local theory: in the limiting case of $\ell\rightarrow 0$ one has $a\rightarrow 1$ such that this quantity vanishes identically. Moreover, since the form factor satisfies $a(0)=1$, the quantity $\Delta\mathcal{G}_{\bar{X}}$ is devoid of any poles in the complex plane and hence analytic. This implies that non-locality, as described in non-local infinite-derivative theories, modifies all local Green functions equally, irrespective of their causal properties.

Concretely, making use of Eq.~\eqref{eq:GFN}, the Green function for our scalar non-local theory takes the form
\begin{align}
\mathcal{G}_{\bar{X}} = \frac{e^{-(\ell^2\bar{\ts{X}}^2)^N}}{-\bar{\ts{X}}^2} \, .
\end{align}
Because the non-local modification does not change the structure of the poles in the complex momentum plane, one might be tempted to perform a similar $i\epsilon$-prescription and contour integration in analogy to the local case. This, however, is impossible, since contour integration assumes a fall-off behavior of the momentum space representation of the Green function which is not satisfied in our non-local infinite-derivative model due to the exponential factor. Incidentally, this problem is well known in the non-local literature and lies at the heart of unitarity issues of non-local theories \cite{Shapiro:2015uxa,Carone:2016eyp,Briscese:2018oyx,Buoninfante:2018mre}.

At this point we note that it is possible to avoid the notion of contour integration by following the approach proposed in Ref.~\cite{Frolov:2016xhq}. Using the Sokhotski--Plemelj theorem for continuous functions it is shown that one may derive non-local Green functions with the correct causal properties by performing a one-dimensional line integral along the real axis. To obtain the retarded Green function we shift the poles by an infinitesimal quantity ${-i\epsilon}$ in accordance to the local theory, and define
\begin{align}
\label{eq:retarded-gf}
\mathcal{G}^\text{R}_{\bar{X}} &\equiv \frac{e^{-(\ell^2\mathrm{\bar{\bs{X}}}^2)^N}}{-\bar{\mathrm{\bs{X}}}^2\big|_{-i\epsilon}}\equiv\frac{e^{-[\ell^2(-\bar{t}^2+\bar{x}^2+\bar{y}^2+\bar{z}^2)]^N}}{-[-(\bar{t}-i\epsilon)^2+\bar{x}^2+\bar{y}^2+\bar{z}^2]} \nonumber \\
&=\frac{e^{-[\ell^2(\sigma_{\bar{u}}\sigma_{\bar{v}}\bar{\zeta}^2+\bar{\rho}^2)]^N}}{-\big[\sigma_{\bar{u}}\sigma_{\bar{v}}\bar{\zeta}^2+\bar{\rho}^2+i\bar{\zeta}\big(\sigma_{\bar{u}}e^{\bar{\tau}}{-}\sigma_{\bar{v}}e^{-\bar{\tau}})\epsilon \big]} \, ,
\end{align}
where in the second line we employed Rindler coordinates that are ideally suited for analytical calculations with uniformly accelerated sources.

To that end, the momentum space description of the source $j_{\bar{X}}$ takes the following form in Rindler coordinates:
\begin{align}
\label{eq:sourcefourier}
j_{\bar{X}} &=\frac{\mu\alpha}{2\pi^2}\int\limits_{M} \dd t \dd x \dd y\dd z\, e^{i(-\bar{t}t+\bar{x}x+\bar{y}y+\bar{z}z)} \nonumber \\
&\hspace{45pt}\times\delta^{(2)}(-t^2+z^2-\alpha^2)\delta(x)\delta(y)\theta(u)\theta(v) \nonumber \\
&=\frac{\mu\alpha}{2\pi^2}\int\limits_{M_{\textrm{R}}} \dd \tau \dd \zeta \, \zeta \exp{\Big[i\tfrac{\bar{\zeta}\zeta}{2}(\sigma_{\bar{u}} e^{\bar{\tau}-\tau}+\sigma_{\bar{v}} e^{-\bar{\tau}+\tau})\Big]} \nonumber \\
&\hspace{45pt}\times \exp\left[ i\left(\bar{x}x+\bar{y}y\right) \right] \delta{}^{(2)}(\zeta^2-\alpha^2) \nonumber \\
&=\frac{\mu\alpha}{4\pi^2}\int\limits_{\mathbb{R}} \dd \tau \, \exp{\left[i\tfrac{\alpha\bar{\zeta}}{2}\left(\sigma_{\bar{u}} e^{\bar{\tau}-\tau}+\sigma_{\bar{v}} e^{-\bar{\tau}+\tau}\right)\right]} \, .
\end{align}
With the expressions for both $\mathcal{G}^\text{R}_{\bar{X}}$ and $j_{\bar{X}}$ known in momentum space we may now utilize Eq.~\eqref{eq:retsolcongr} to arrive at the real-space expression of the retarded field $\phi$. Since we shall employ Rindler coordinates, this step involves the integration over four distinct patches of momentum space, which we refer to as $M_{\bar{\textrm{W}}}$ (with $\bar{\textrm{W}} = \bar{\textrm{R}},\bar{\textrm{L}},\bar{\textrm{F}},\bar{\textrm{P}}$ in analogy to the real-space covering of Minkowski spacetime). For this reason the integration can be split in four integrals $I_{\bar{\textrm{W}}}$ over the regions ${M_{\bar{\textrm{W}}}}$. The retarded solution for $\phi$ is then given by four contributions,
\begin{align}
\phi(X) &= \sum\limits_{\bar{\textrm{W}}} I_{\bar{\textrm{W}}}(X) \, , \\
I_{\bar{\textrm{W}}}(X) &= \frac{1}{4\pi^2}\int\limits_{M_{\bar{\textrm{W}}}}\!\mathfrak{g}^{1/2}(\bar{X}) e^{-i\bar{\bs{X}}\cdot\bs{X}} ~ \mathcal{G}^\text{R}_{\bar{X}} \, j_{\bar{X}} \, .
\end{align}
Then, employing the integral expression for the source as per Eq.~\eqref{eq:sourcefourier}, $I_{\bar{\textrm{W}}}$ takes the following rather lengthy form:
\begin{widetext}
\begin{align}
    \label{eq:main-calc}
    I_{\bar{\textrm{W}}}(X)
    &=\frac{\mu\alpha}{8\pi^3}\!\int\limits_{\mathbb{R}}\! \dd\bar{\tau}\int\limits_0^\infty\! \dd\bar{\zeta}\!\int\limits_{\mathbb{R}} \!\dd\tilde{\tau}\,\bar{\zeta}\exp\left\{i\tfrac{\bar{\zeta}}{2}\left[\sigma_{\bar{u}}(\alpha e^{-\tilde{\tau}}{-}\sigma_v \zeta e^{-\tau})e^{\bar{\tau}}{+}\sigma_{\bar{v}}(\alpha e^{\tilde{\tau}}{-}\sigma_u \zeta e^{\tau})e^{-\bar{\tau}}\right]\right\} \nonumber
    \\
    &\feq\times\bigg[\fint\limits_0^{\infty} \!\dd\bar{\rho}\, \bar{\rho}J_0(\rho\bar{\rho})\frac{e^{-[\ell^2(\sigma_{\bar{u}}\sigma_{\bar{v}}\bar{\zeta}^2+\bar{\rho}^2)]^N}}{-(\sigma_{\bar{u}}\sigma_{\bar{v}}\bar{\zeta}^2+\bar{\rho}^2)}
    +\frac{i\pi(\sigma_{\bar{u}}{-}\sigma_{\bar{v}})}{2}\bar{\zeta}J_0(\rho\bar{\zeta})\int\limits_0^{\infty} \!\dd\bar{\rho}\,\delta{}^{(2)}(\sigma_{\bar{u}}\sigma_{\bar{v}}\bar{\zeta}^2+\bar{\rho}^2)\bigg]
    \\
    &=\frac{\mu\alpha}{8\pi^3}\!\int\limits_{\mathbb{R}}\! \dd\bar{\tau}\int\limits_0^\infty\! \dd\bar{\zeta}\!\int\limits_{\mathbb{R}} \!\dd\tilde{\tau}\,\bar{\zeta}\exp\left\{i\tfrac{\bar{\zeta}}{2}\left[\sigma_{\bar{u}}(\alpha e^{-\tilde{\tau}}{-}\sigma_v \zeta e^{\bar{\tau}}){+}\sigma_{\bar{v}}(\alpha e^{\tilde{\tau}}{-}\sigma_u \zeta e^{-\bar{\tau}})\right]\right\}\! \nonumber
    \\
    &\feq\times\bigg[\fint\limits_0^{\infty} \!\dd\bar{\rho}\, \bar{\rho}J_0(\rho\bar{\rho})\frac{e^{-[\ell^2(\sigma_{\bar{u}}\sigma_{\bar{v}}\bar{\zeta}^2+\bar{\rho}^2)]^N}}{-(\sigma_{\bar{u}}\sigma_{\bar{v}}\bar{\zeta}^2+\bar{\rho}^2)}
   +\frac{i\pi(\sigma_{\bar{u}}{-}\sigma_{\bar{v}})}{4}J_0(\rho\bar{\zeta})\bigg]\;.
\end{align}
\end{widetext}
In the above we first integrated out the angles,
\begin{equation}
    \int\limits_0^{2\pi}\!\dd\bar{\varphi} \, \exp\left[-i\bar{\rho}\rho\cos{(\bar{\varphi}-\varphi)} \right]=2\pi J_0(\bar{\rho}\rho) \, ,
\end{equation}
where $J_0(x)$ denotes the Bessel function of the first kind \cite{Olver:2010}. Then we made use of the Sokhotski--Plemelj theorem to rewrite the regulated expression
\begin{align}
\begin{split}
\frac{f(\bar{\rho})}{-\bar{\mathrm{\bs{X}}}^2\big|_{-i\epsilon}} &= \textrm{p.v.}_{\bar{\rho}}\frac{f(\bar{\rho})}{-\bar{\mathrm{\bs{X}}}^2} \label{eq:sokplem} \\
&\hspace{12pt} + \frac{i\pi(\sigma_{\bar{u}}{-}\sigma_{\bar{v}})}{2}f(\bar{\zeta})\delta{}^{(2)}(\bar{\mathrm{\bs{X}}}^2)\;,
\end{split}
\end{align}
where $f(\bar{\rho})$ is a continuous function. Due to the central importance for the causal properties of the solution presented in this paper, we prove the above relation in detail in Appendix~\ref{sc:SPtheorem}.

In the above, the symbol $\textrm{p.v.}_{\bar{\rho}}$ denotes the Cauchy principal value with respect to the variable $\bar{\rho}$ with other coordinates held fixed. The symbol $\fint$ denotes that the integration is to be performed with the standard prescription for the Cauchy principal value. Note that the last term of \eqref{eq:sokplem}, including the $\delta$-distribution, has support only in $M_{\bar{\textrm{F}}}\cup M_{\bar{\textrm{P}}}$ as there are no poles in $M_{\bar{\textrm{R}}}\cup M_{\bar{\textrm{L}}}$. Consequently, in the regions $M_{\bar{\textrm{R}}}\cup M_{\bar{\textrm{L}}}$ the Cauchy principal value integral reduces to the standard integral and \eqref{eq:sokplem} yields the identity, as it must. 

Then, in the second equality of Eq.~\eqref{eq:main-calc}, we integrated out the $\delta$-distribution and shifted the variables $\tilde{\tau}$ and $\bar{\tau}$. In order to obtain the final expression for the retarded field $\phi$ we now need to sum the contributions $I_{\bar{W}}$, and it is useful to first sum the integrals corresponding to the opposite regions. We arrive at the compact expressions
\begin{align}
    I^{\pm}(X) &\equiv \left\{
    \begin{aligned} I_{\bar{\textrm{R}}}(X) + I_{\bar{\textrm{L}}}(X)
    \\[1pt]
    I_{\bar{\textrm{F}}}(X) + I_{\bar{\textrm{P}}}(X)
    \end{aligned}\right . \nonumber \\
    &=\frac{\mu\alpha}{4\pi^3}\int\limits_0^\infty\! \dd\bar{\zeta}\,\bar{\zeta}\bigg[C^{\pm}_{\textrm{W}}(\zeta,\bar{\zeta})\fint\limits_0^\infty \!\dd\bar{\rho}\, \bar{\rho}J_0(\rho\bar{\rho})\frac{e^{-[\ell^2(\pm\bar{\zeta}^2+\bar{\rho}^2)]^N}}{-(\pm\bar{\zeta}^2+\bar{\rho}^2)} \nonumber \\
    &\hspace{20pt} -\frac{\pi}{2}\theta_\mp S_{\textrm{W}}(\zeta,\bar{\zeta}) J_0(\rho\bar{\zeta})\bigg]\;, \label{eq:ipim}
\end{align}
where we defined $\theta_+ = 1$, $\theta_-=0$, and $C^{\pm}_{\textrm{W}}(\zeta,\bar{\zeta})$ and $S_{\textrm{W}}(\zeta,\bar{\zeta})$ denote the following cosine and sine integrals:
\begin{equation}
\begin{aligned}
    C^{\pm}_{\textrm{W}}(\zeta,\bar{\zeta}) &\equiv\int\limits_{\mathbb{R}}\! \dd\bar{\tau}\!\int\limits_{\mathbb{R}} \!\dd\tilde{\tau}\cos\left[\tfrac{\bar{\zeta}}{2}(\alpha e^{-\tilde{\tau}}{-}\sigma_v \zeta e^{\bar{\tau}}) \right. \\
    &\hspace{20pt} \pm \left. \tfrac{\bar{\zeta}}{2}(\alpha e^{\tilde{\tau}}{-}\sigma_u \zeta e^{-\bar{\tau}})\right] \;, \\
    S_{\textrm{W}}(\zeta,\bar{\zeta}) &\equiv\int\limits_{\mathbb{R}}\! \dd\bar{\tau}\!\int\limits_{\mathbb{R}} \!\dd\tilde{\tau}\sin\left[\tfrac{\bar{\zeta}}{2}(\alpha e^{-\tilde{\tau}}{-}\sigma_v \zeta e^{\bar{\tau}}) \right. \\
    &\hspace{20pt} - \left. \tfrac{\bar{\zeta}}{2}(\alpha e^{\tilde{\tau}}{-}\sigma_u \zeta e^{-\bar{\tau}})\right]\;.
\end{aligned}
\end{equation}
These double integrals can be separated into products of integrals (see Eq.~(3.868), (1)--(4) in Ref.~\cite{gradshteyn1996table}) and take the following form in the various regions of Minkowski spacetime:
\begin{align}
    C^{\pm}_{\textrm{R}}(\zeta,\bar{\zeta}) &=
    \begin{cases}
    \pi ^2 \left[ J_0(\alpha\bar{\zeta}) J_0(\zeta \bar{\zeta})+Y_0(\alpha\bar{\zeta}) Y_0(\zeta \bar{\zeta}) \right]\;,
    \\[5pt]
    4 K_0(\alpha\bar{\zeta}) K_0(\zeta \bar{\zeta})\;,
    \end{cases} \nonumber \\
    C^{\pm}_{\textrm{L}}(\zeta,\bar{\zeta}) &=
    \begin{cases}
    \pi ^2 \left[ -J_0(\alpha\bar{\zeta}) J_0(\zeta \bar{\zeta})+Y_0(\alpha\bar{\zeta}) Y_0(\zeta \bar{\zeta}) \right]\;,
    \\[5pt]
    4 K_0(\alpha\bar{\zeta}) K_0(\zeta \bar{\zeta})\;,
    \end{cases} \\ \nonumber
    C^{\pm}_{\textrm{F}}(\zeta,\bar{\zeta}) &=C^{\pm}_{\textrm{P}}(\zeta,\bar{\zeta})=
    \begin{cases}
    -2 \pi  Y_0(\alpha\bar{\zeta}) K_0(\zeta \bar{\zeta})\;,
    \\[5pt]
    -2 \pi  K_0(\alpha\bar{\zeta}) Y_0(\zeta \bar{\zeta})\;,
    \end{cases} \\  \nonumber
    S_{\textrm{R}}(\zeta,\bar{\zeta}) &=S_{\textrm{L}}(\zeta,\bar{\zeta})=0\;,\\
    S_{\textrm{F}}(\zeta,\bar{\zeta}) &= -S_{\textrm{P}}(\zeta,\bar{\zeta}) = 2\pi K_0(\alpha\bar{\zeta}) J_0(\zeta \bar{\zeta}) \; .  \nonumber 
\end{align}
Then, the final solution for $\phi$ is given by the sum
\begin{equation}\label{eq:phiIpIm}
\phi(X) = I^{+}(X)+I^{-}(X)\;.
\end{equation}
We were not able to find a closed-form analytic expression for $\phi$, which is why we refrain from giving an explicit expression at this point.

\subsection{Local case ${\ell=0}$}
\label{sec:local-case}
As a simple consistency check let us recover the known local solution for $\ell\rightarrow 0$. Employing Eq.~\eqref{eq:ipim} we find
\begin{align}
\phi(X) &=\frac{\mu\alpha}{4\pi^3}\int\limits_0^\infty \!\dd\bar{\zeta}\, \bar{\zeta}\bigg[{-}  K_0(\rho \bar{\zeta})C^{+}_{\textrm{W}}(\zeta,\bar{\zeta}) \label{eq:philocalintegral} \\
&\hspace{20pt}+\frac{\pi}{2} Y_0(\rho \bar{\zeta})C^{-}_{\textrm{W}}(\zeta,\bar{\zeta})-\frac{\pi}{2} J_0(\rho\bar{\zeta})S_{\textrm{W}}(\zeta,\bar{\zeta})\bigg] \;, \nonumber
\end{align}
where we used the following principal value integral expressions (for $\rho \not= 0$):
\begin{equation}
    \fint\limits_0^{\infty} \!\dd\bar{\rho}\, \frac{\bar{\rho}J_0(\rho\bar{\rho})}{-(\pm\bar{\zeta}^2+\bar{\rho}^2)}=
    \begin{cases}
        -K_0(\rho \bar{\zeta})\;,
        \\[5pt]
        \frac{\pi}{2}Y_0(\rho \bar{\zeta})\;.
    \end{cases}
\end{equation}
Numerical integration of \eqref{eq:philocalintegral} perfectly matches the known analytic result for the retarded solution \cite{Zelnikov:1982,Ren:1993bs,Bicak:2002yk,Bicak:2005yt},
\begin{equation}
\label{eq:locretsol}
\phi(X) = -\frac{\mu\alpha}{4\pi}\frac{\theta{(u)}}{\sqrt{(\sigma_v \zeta^2+\rho^2+\alpha^2)^2/4-\sigma_v \alpha^2 \zeta^2}} \, .
\end{equation}
Note that this field is non-zero in ${M_{\textrm{R}}\cup M_{\textrm{F}}}$ and vanishes in ${M_{\textrm{L}}\cup M_{\textrm{P}}}$. Despite the discontinuity across the surface ${u=0}$, it fully satisfies the field equations with distributional source \eqref{eq:source}. The advanced solution $\phi_{\textrm{A}}$ can be found by formally reversing the time direction, ${t\to{-}t}$, which is equivalent to the exchange ${u\leftrightarrow v}$,
\begin{equation}
\phi^{\text{A}}(X) = -\frac{\mu\alpha}{4\pi}\frac{\theta(v)}{\sqrt{(\sigma_u \zeta^2+\rho^2+\alpha^2)^2/4-\sigma_u \alpha^2 \zeta^2}} \, .
\end{equation}
As already pointed out in the Introduction, this local solution is singular at the location of the source, that is, in the plane $\zeta=\alpha$ whenever $\rho=0$. On the future horizon $t=z$, however, the retarded field is regular. For a more detailed discussion of this local solution, including quantum radiation, we refer to Ren and Weinberg \cite{Ren:1993bs}.

\subsection{Non-local case ${\ell>0}$}
Let us now study the non-local case ${\ell>0}$. In general, for $\rho >0$, we were not able to proceed analytically with the integral expressions for the non-local retarded field via Eqs.~\eqref{eq:main-calc}, \eqref{eq:ipim}, and \eqref{eq:phiIpIm}. Restricting ourselves to the plane ${\rho=0}$, however, the solution reduces to
\begin{align}
    \phi_0(X) &\equiv \phi(X)|_{\rho=0} \nonumber \\
    &=\frac{\mu\alpha}{4 \pi ^3}\int\limits_0^\infty\!\dd\bar{\zeta}\,\bar{\zeta} \bigg[\frac{\Ei\big({-}\ell^{2N}\bar{\zeta}^{2N}\big)}{2N}C^{+}_{\textrm{W}}(\zeta,\bar{\zeta}) \label{eq:nonlocretsol} \\
    &\hspace{12pt} +\frac{\Ei\big({-}({-}\ell^2)^{N}\bar{\zeta}^{2N}\big)}{2N}C^{-}_{\textrm{W}}(\zeta,\bar{\zeta}) -\frac{\pi}{2} S_{\textrm{W}}(\zeta,\bar{\zeta})\bigg]\;, \nonumber
\end{align}
where we used the principal value integral expression
\begin{equation}
\label{eq:non-local-pv}
    \fint\limits_0^{\infty} \!\dd\bar{\rho}\, \frac{\bar{\rho}\, e^{-\left[\ell^2(\sigma_{\bar{u}}\sigma_{\bar{v}}\bar{\zeta}^2+\bar{\rho}^2)\right]^N}}{-\big(\pm\bar{\zeta}^2+\bar{\rho}^2\big)}=
        \frac{\Ei\big({-}(\pm \ell^2)^{N} \bar{\zeta}^{2N}\big)}{2N}\;,
\end{equation}
and $\text{Ei}(x)$ denotes the exponential integral \cite{Olver:2010}. Inspecting \eqref{eq:nonlocretsol} one immediately notices that the cases of even and odd $N$ are quite different. Indeed, the integrals for odd values of $N$ do not converge. This can be shown simply for $X\in M_{\textrm{R}}\cup M_{\textrm{L}}$. In this case, the first and third term in the integrand of \eqref{eq:nonlocretsol} are suppressed for large values of $\bar{\zeta}$, but the second term grows to infinity. For $X\in M_{\textrm{F}}\cup M_{\textrm{P}}$, the second term is also unbounded because it oscillates with growing amplitude. On the other hand, the integral converges for even values of $N$. This behaviour for even/odd non-local theories seems to be in agreement with Ref.~\cite{Frolov:2016xhq,Boos:2019fbu}.

For numerical analysis it is useful to introduce dimensionless quantities. Since we assume the scale of non-locality $\ell$ to be fundamental, we choose to normalize the physical parameters of distance and acceleration with respect to that length scale and introduce the quantity
\begin{align}
\hat{\alpha} \equiv \frac{\alpha}{\ell} \, .
\end{align}
The scalar field is proportional to the mass of the particle $\mu$. Since that constant does not appear anywhere else we define the dimensionless scalar field $\hat{\phi}$ as
\begin{align}
\hat{\phi} \equiv \frac{\phi \ell}{\mu} \, .
\end{align}
Now the only free parameter is the dimensionless acceleration parameter $\hat{\alpha}$, which measures inverse acceleration per unit mass.

\begin{figure*}
    \centering
    \includegraphics[width=0.8\textwidth]{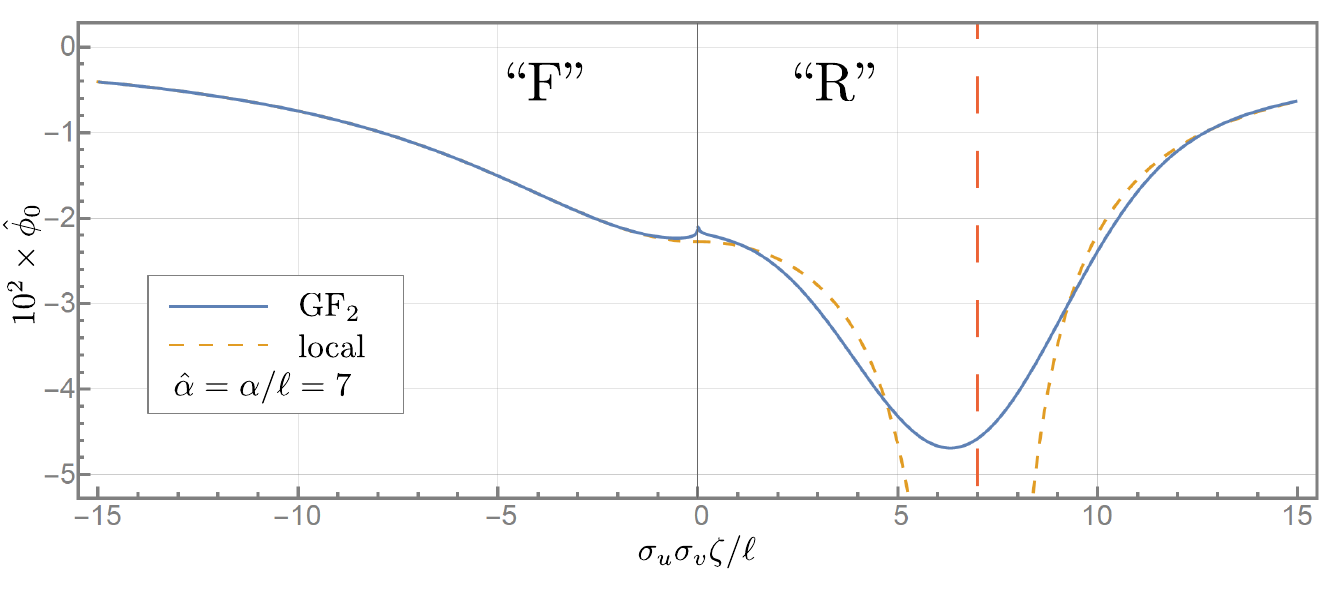}
    \includegraphics[width=0.8\textwidth]{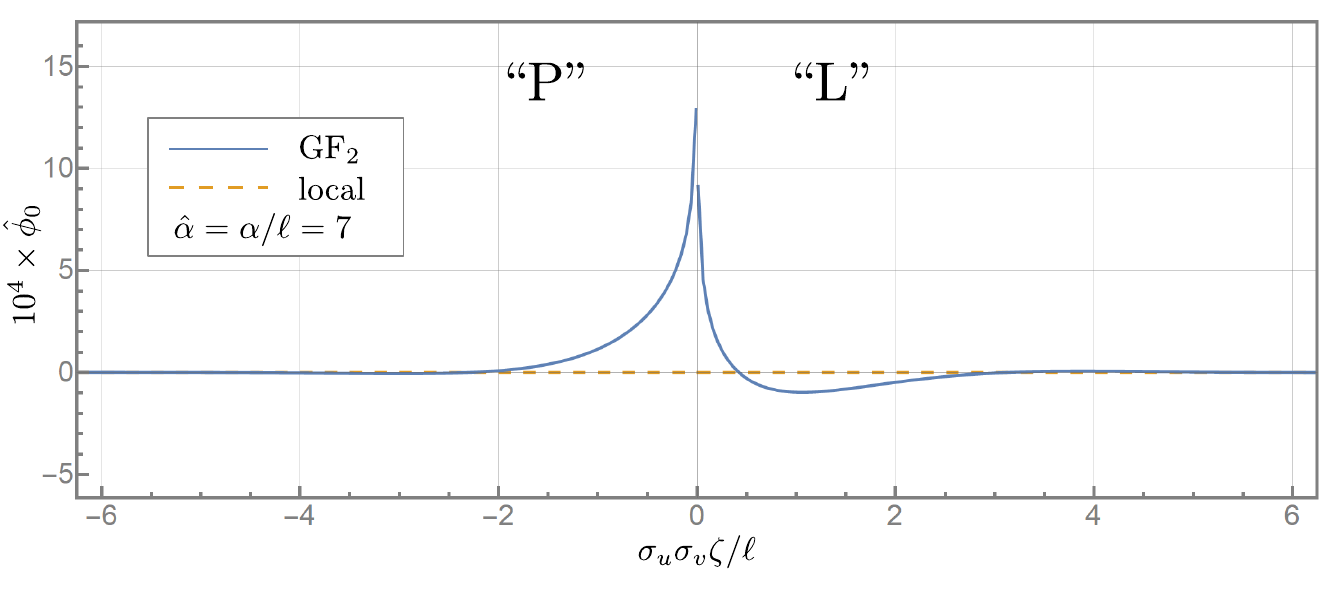}
    \caption{The local (dashed) and non-local (solid) retarded dimensionless field in the four Rindler wedges plotted as a function of $\sigma_u\sigma_v\zeta$, in the plane $\rho=0$ for a dimensionless acceleration of $\hat{\alpha} = 7$, with a dimensionless step size of $0.05$. The vertical dashed line in the right wedge indicates the position of the particle at $\hat{\zeta} = \hat{\alpha}$.}
    \label{fig:solutionN2}
\end{figure*}

For the remainder of this paper let us focus on the simplest case of $N=2$, which we refer to as $\mathrm{GF_2}$ theory. Then one finds
\begin{align}
\hat{\phi}_0(\hat{X}) &=\frac{\hat{\alpha}}{4 \pi ^3}\int\limits_0^\infty\!\dd\bar{\zeta}\,\bar{\zeta} \bigg[\frac{\Ei\left(-\bar{\zeta}^4\right)}{4}C^{+}_{\textrm{W}}(\hat{\zeta},\bar{\zeta}) \label{eq:nonlocretsol-n=2} \\
    &\hspace{12pt} +\frac{\Ei\left(-\bar{\zeta}^4\right)}{4}C^{-}_{\textrm{W}}(\hat{\zeta},\bar{\zeta}) -\frac{\pi}{2} S_{\textrm{W}}(\hat{\zeta},\bar{\zeta})\bigg]\;, \nonumber
\end{align}
where we introduced the dimensionless distance $\hat{\zeta} \equiv \zeta/\ell$. The integration can be performed numerically for each Rindler wedge, and we plot a graphical representation in Fig.~\ref{fig:solutionN2}. For convenience we combined the numerical expressions for the right and future wedge by artificially plotting $\phi$ as a function of $\sigma_u\sigma_v\zeta$, and, similarly, in the left and past Rindler wedge.

The retarded field has several noteworthy properties:
\begin{itemize}
\item[1.] For large timelike and spacelike distances, $\zeta\gg\ell$, one recovers the local result discussed in the previous section.\\[-1.4\baselineskip]
\item[2.] The non-local field is regular at the location of the source, $\zeta=\alpha$, in contrast to the local field.\\[-1.4\baselineskip]
\item[3.] The non-local field is non-vanishing in the left and past Rindler wedge, unlike the local field.\\[-1.4\baselineskip]
\item[4.] The behavior of the non-local solution around the horizon appears singular. Closer inspection, as we shall discuss below, reveals that this is an artefact of the unphysical assumption of uniform acceleration.
\end{itemize}
Let us now discuss these properties of the retarded non-local field in more detail.

\subsubsection{Asymptotic timelike and spacelike behavior}

As discussed in Sec.~\ref{sec:local-case}, in the local limit $\ell\rightarrow 0$ one recovers the local expression for the retarded field. However, we may also consider the dimensionless limit $\hat{\zeta} \equiv \zeta/\ell\rightarrow \infty$, which corresponds to the limit of $\ell\rightarrow 0$ at finite Rindler radius $\zeta$, or to the large-distance limit in the case of finite $\ell>0$. From the graphical representation in Fig.~\ref{fig:solutionN2} it is clear that the non-local retarded field approaches the values of the local theory at large spacelike and timelike distances,
\begin{align}
\hat{\phi}_0(\hat{\zeta} \gg 1) = -\frac{\hat{\alpha}(1+\sigma_u)}{4\pi\hat{\zeta}^2} \, .
\end{align}
This is a non-trivial consistency check since it implies that for large timelike and spacelike distances the effects of non-locality are heavily suppressed.

\subsubsection{Regularity at the location of the source}

In stark contrast to the local solution \eqref{eq:locretsol}, the non-local field is \emph{finite} at the location of the particle, $\zeta=\alpha$. It is possible to calculate this value analytically,
\begin{align}
\hat{\phi}_0(\hat{\zeta} \approx \hat{\alpha}) &= \frac{1}{64\pi^{7/2}\hat{\alpha}} \left[ -4\pi^3\hat{\alpha}^2 {}_2 F{}_4\left( \tfrac14,\tfrac34; ~\tfrac12,1,1,\tfrac32; \frac{\hat{\alpha}^4}{16} \right) \right. \nonumber \\
&+\pi^{5/2}\hat{\alpha}^4 {}_2 F{}_4\left( \tfrac34,\tfrac54; ~ \tfrac32,\tfrac32,\tfrac32,2; \frac{\hat{\alpha}^4}{16} \right) \nonumber \\
&-\sqrt{2}\pi^2\hat{\alpha}^2 A(\hat{\alpha}) -2\sqrt{2} B(\hat{\alpha}) \Big] \nonumber \\
&\hspace{11pt}+ \mathcal{O}(\hat{\zeta}-\hat{\alpha}) \, , \\
A(\hat{\alpha}) &\equiv G{}^{62}_{69}\left( \left. \begin{matrix*}[l]\tfrac12,\tfrac12;~\tfrac14,\tfrac14,\tfrac34,\tfrac34 \\[3pt] 0,0,0,\tfrac12,\tfrac12,\tfrac12; ~ -\tfrac12, \tfrac14, \tfrac34 \end{matrix*}\right| \frac{\hat{\alpha}^4}{16} \right) \, , \\
B(\hat{\alpha}) &\equiv G{}^{62}_{47}\left( \left. \begin{matrix*}[l] 1,1;~ \tfrac34,\tfrac54 \\[3pt] \tfrac12,\tfrac12,\tfrac12,1,1,1;~ 0 \end{matrix*}\right| \frac{\hat{\alpha}^4}{16} \right) \, , 
\end{align}
where $G^{mn}_{pq}$ denote Meijer G-functions \cite{Olver:2010}. One may show that $\hat{\phi}(\hat{\zeta} \approx \hat{\alpha})$ is finite, smooth, and negative for positive values of $\hat{\alpha}$, whereas it vanishes for $\hat{\alpha}\rightarrow 0$. This manifestly finite behavior at the location of the source matches our expectation that non-locality regularizes the field of localized sources and, perhaps more importantly, presents a concrete extension from previous static and stationary results known in the literature to the full, time-dependent case.

A closer inspection reveals that the linear term $\mathcal{O}(\hat{\zeta}-\hat{\alpha})$ does not vanish. This corresponds to the fact that the minimum of the non-local potential is not located at $\zeta=\alpha$ but, rather, is shifted towards smaller values of $\zeta$. This behavior can also be seen in Fig.~\ref{fig:solutionN2}.

\subsubsection{Causal properties}

Recall that the local solution \eqref{eq:locretsol} is proportional to the step function $\theta(u)$, implying that the local retarded field is strictly zero in the left and past Rindler wedges. In the non-local case one might expect that this is no longer the case. And indeed, Fig.~\ref{fig:solutionN2} confirms this suspicion: the non-local retarded field is non-zero in the left and past Rindler wedges. While we were unable to find a complete analytical description, it is again possible to find the value of the field analytically at the somewhat ad hoc location $\zeta=\alpha$. In the left Rindler wedge we find
\begin{align}
\begin{split}
\hat{\phi}_0(\hat{\zeta} \approx \hat{\alpha}) &= -\frac{1}{32\sqrt{2}\pi^{7/2}\hat{\alpha}} \left[ \pi^2\hat{\alpha}^2 A(\hat{\alpha}) +2 B(\hat{\alpha}) \right] \\
&\hspace{11pt}+ \mathcal{O}(\hat{\zeta}-\hat{\alpha}) \, ,
\end{split}
\end{align}
whereas for the past Rindler wedge one has
\begin{align}
\begin{split}
\hat{\phi}_0(\hat{\zeta} \approx \hat{\alpha}) &= \frac{1}{8\pi\hat{\alpha}} - \frac{\hat{\alpha}}{64\pi^{5/2}} G{}^{31}_{14}\left( \left. \begin{matrix*}[l]\tfrac12,; \\[3pt] 0,0,0; ~ -\tfrac12 \end{matrix*}\right| \frac{\hat{\alpha}^4}{64} \right) \\
&\hspace{11pt}+ \mathcal{O}(\hat{\zeta}-\hat{\alpha}) \, ,
\end{split}
\end{align}
Let us emphasize that these non-zero values arise solely due to the presence of non-locality, $\ell>0$. In the limit of vanishing non-locality and finite acceleration parameter $\alpha$ one has $\hat{\alpha} = \alpha/\ell \rightarrow \infty$, and one may show that in this limit the above terms vanish identically.

In linearized non-local theories it is common wisdom that ``non-locality smears out sharp sources'' \cite{Giacchini:2018wlf,Boos:2020qgg}, and one might be tempted to interpret the above expressions as the result of a smeared out step function similar to the expression $\sim e^{\ell^2\Box^2} \theta(u)$. However, due to the lack of concrete analytical expressions for the retarded field for arbitrary values of $\zeta$ it is not possible to test this idea further.

\subsubsection{Singular behavior in vicinity of acceleration horizons}

From our numerical plot in Fig.~\ref{fig:solutionN2} it is obvious that the retarded non-local field behaves somewhat singularly in proximity to the acceleration horizons. Expanding the integrand \eqref{eq:nonlocretsol} for small values of $\hat{\zeta}$ one finds the following logarithmic behavior:
\begin{align}
\hat{\phi}_0(\hat{\zeta} \ll 1) &= c_0(\hat{\alpha}) + c_1(\hat{\alpha}) \log \hat{\zeta} + \mathcal{O}(\hat{\zeta}^2) \, , \label{eq:div-1} \\
c_0(\hat{\alpha}) &= \int\limits_{\mathbb{R}} \dd\bar{\zeta} \frac{\hat{\alpha}\bar{\zeta}}{32\pi^3} \Big\{ \pi\text{Ei}(-\bar{\zeta}^4)\Big[ \pi(\sigma_u+\sigma_v)J_0(\hat{\alpha}\bar{\zeta}) \nonumber \\
&+ 4Y_0(\hat{\alpha}\bar{\zeta})\left(\gamma + \log\tfrac{\bar{\zeta}}{2} \right) \Big] \nonumber \\
&- 4K_0(\hat{\alpha}\bar{\zeta})\Big[ \pi^2 (\sigma_u-\sigma_v) \nonumber \\
&+2\text{Ei}(-\bar{\zeta}^4) \left(\gamma + \log\tfrac{\bar{\zeta}}{2} \right) \Big] \Big\} \, , \label{eq:c0-constant} \\
c_1(\hat{\alpha}) &= \frac{1}{4\pi^4\hat{\alpha}} G^{42}_{25}\left( \left. \begin{matrix*}[l]1,1; \\[3pt] \tfrac12, \tfrac12, 1, 1 ;~ 0 \end{matrix*}\right| \frac{\hat{\alpha}^4}{256} \right) \nonumber \\
& -\frac{1}{2\pi^2\hat{\alpha}} G{}^{42}_{47}\left( \left. \begin{matrix*}[l]1,1; ~ \tfrac14,\tfrac34 \\[3pt] \tfrac12, \tfrac12, 1, 1;~ 0,\tfrac14,\tfrac34\end{matrix*}\right| \frac{\hat{\alpha}^4}{256} \right) \, . \label{eq:c1-constant}
\end{align}
Apparently, the retarded field diverges logarithmically as one approaches the acceleration horizon. Note that $c_0$ depending on $\sigma_u$ and $\sigma_v$ leads to different values in different Rindler wedges. However, the constant $c_1$ that multiplies the diverging logarithmic term is universal.

This logarithmic divergence arises due to non-locality and is pathological as the retarded field of the local theory does not exhibit any singular behavior, except for a discontinuity on the past horizon, which we shall address in the next subsection. In what follows we will demonstrate that the pathological logarithmic divergence arises solely due to the unphysical assumption of a uniformly accelerated massive particle. This acceleration would require an infinite amount of energy and result in a massive particle moving asymptotically at the speed of light.

\begin{figure}
    \centering
    \includegraphics[width=0.45\textwidth]{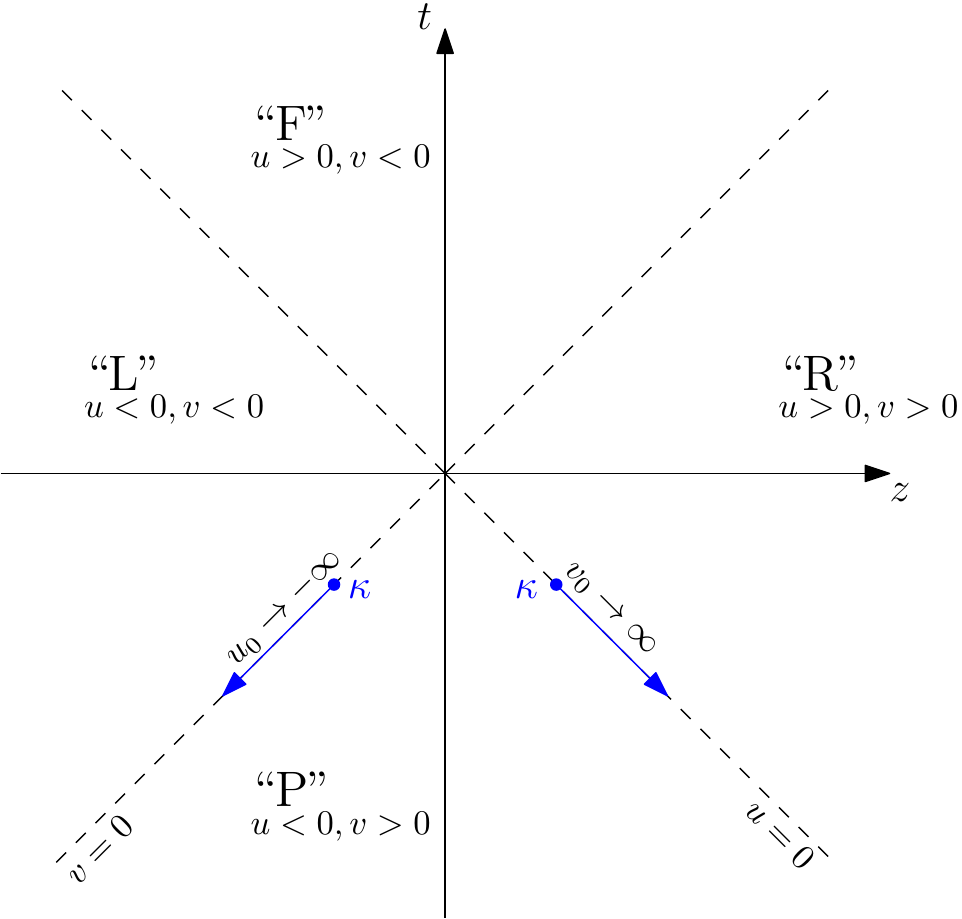}
    \caption{Test setup to understand the emergence of singular behavior on the past and future acceleration horizon due to non-locality. Consider two sources of magnitude $\kappa$ located at $(u_0,0)$ and $(0,v_0)$ and then take the limit $u_0\rightarrow-\infty$ and $v_0\rightarrow\infty$, shifting the sources into the asymptotic past.}
    \label{fig:sources}
\end{figure}

In order to gain some qualitative understanding of the divergences, let us consider a simpler setting of a single point-like source located at $(u_0, v_0)$,
\begin{align}
\label{eq:test-source}
j_\text{test}(X) = \kappa \delta(u-u_0)\delta(v-v_0) \delta(x)\delta(y) \, ,
\end{align}
where $\kappa$ is a dimensionless prefactor. Focusing our considerations to the plane $\rho=0$ the resulting field is then simply
\begin{align}
\phi_{\text{test}}(X) = \kappa \mathcal{G}(u,v;u_0,v_0) \, .
\end{align}
To study the effects of non-locality it is sufficient to consider the non-local modification of the Green function. Since we are interested in a source that becomes asymptotically null we need to check two cases:
\begin{itemize}
\item[(a)] Past horizon: Set ${u_0=0}$ and consider the resulting field in the limit ${v_0\rightarrow\infty}$, evaluated on the past horizon ($u=0$).\\[-1.4\baselineskip]
\item[(b)] Future horizon: Set ${v_0=0}$ and consider the resulting field in the limit ${u_0\rightarrow-\infty}$, evaluated on the future horizon ($v=0$).
\end{itemize}
For a visualization we refer to Fig.~\ref{fig:sources}. The non-local Green function modification---see Appendix~\ref{app:deltaG}---can be written as follows:
\begin{align}
\Delta\mathcal{G} &= \frac{1}{4\pi^{5/2}s^2} \int\limits_0^\infty \dd y e^{-y^2/4} \sin\left( \frac{s^2}{4y\ell^2} \right) \\
&= -\frac{\text{sgn}\left[(u-u_0)(v-v_0)\right]}{4\pi^{5/2}\ell^2} \int\limits_0^\infty \dd y \, \sin\left( \frac{1}{4y} \right) \nonumber \\
&\hspace{20pt} \times \exp\left[ -y^2(u-u_0)^2(v-v_0)^2/(4\ell^4)\right] \, ,
\end{align}
where $s^2=(u-u_0)(v-v_0)$. In order to probe the divergence on the past horizon we set $u_0=0$. If $u \not=0$ then $\Delta\mathcal{G} \equiv 0$ due to the exponential suppression in the limit $v_0\rightarrow\infty$. If $u=0$, then the integral diverges logarithmically close to the past horizon. For the future horizon the analysis goes through identically, \textit{mutatis mutandis}.

An analytic representation of the non-local Green function modification confirms this behavior:
\begin{align}
\Delta\mathcal{G}(s^2) &= \frac{|s^2|}{1024\pi^2\ell^4} G{}^{20}_{03} \left( \left. \begin{matrix} \\ -\tfrac12, -\tfrac12; -1 \end{matrix} \right| \frac{s^4}{256\ell^4} \right) \, ,
\end{align}
where $G{}^{20}_{03}$ denotes the Meijer G-function \cite{Olver:2010}. For a derivation of this expression we refer to Appendix~\ref{app:deltaG}. This function has the following asymptotics:
\begin{align}
\Delta\mathcal{G}(|s^2| \ll 1) &= \frac{1}{32\pi^{5/2}\ell^2} \left[ 2 - 3\gamma -\log\left( \frac{s^4}{64\ell^4} \right) \right] \, , \nonumber \\
\Delta\mathcal{G}(|s^2| \gg 1) &= \frac{1}{2\sqrt{3}\pi^2 |s^2|} \sin\left( \frac{3\sqrt{3}s^{4/3}}{8\cdot2^{2/3}\ell^{4/3}} \right) \\
&\hspace{20pt} \times \exp\left(-\frac{3 s^{4/3}}{8\cdot2^{2/3}\ell^{4/3}}\right) \, . \nonumber
\end{align}
While for large arguments the non-local contributions are strongly suppressed, on the light cone the modifications diverge logarithmically. This means that non-locality may have a non-trivial influence if the point of observation $(u,v)$ and the location of a source at $(u_0,v_0)$ are null separated. This is precisely what happens on the acceleration horizons.

Extracting the prefactor of the logarithmic divergence created by the presence of the test source \eqref{eq:test-source},
\begin{align}
Z_0 \equiv -\frac{\kappa}{8\pi^{5/2}\ell^2} \, ,
\end{align}
we may now equate it to the negative of the near-horizon constant $c_1(\hat{\alpha})$ of Eq.~\eqref{eq:c1-constant} while simultaneously restoring a dimensional $\phi$-field, resulting in an expression for the dimensionless constant $\kappa$,
\begin{align}
\begin{split}
\kappa &= -4\sqrt{\pi}\frac{\mu\ell^2}{\alpha} \left[ G{}^{42}_{47}\left( \left. \begin{matrix*}[l]1,1; ~ \tfrac14,\tfrac34 \\[3pt] \tfrac12, \tfrac12, 1, 1;~ 0,\tfrac14,\tfrac34\end{matrix*}\right| \frac{\alpha^4}{256\ell^4} \right) \right. \\
&\hspace{12pt}- \left. \frac{1}{2\pi^2} G^{42}_{25}\left( \left. \begin{matrix*}[l]1,1; \\[3pt] \tfrac12, \tfrac12, 1, 1 ;~ 0 \end{matrix*}\right| \frac{\alpha^4}{256\ell^4} \right) \right] \, .
\end{split}
\end{align}
This implies that it is possible to regularize the logarithmic divergence by adding a counterterm-like source with the above prefactor on the past horizons at both $(u_0\rightarrow-\infty,v_0=0)$ and $(u_0=0,v_0\rightarrow\infty)$; see also Fig.~\ref{fig:sources}. It is clear that this procedure is necessitated solely due to the presence of non-locality, since in the limit $\ell\rightarrow 0$ one has $\kappa \rightarrow 0$, as expected.

Hence, just as in the local case, the singular behavior arises due to the unphysical assumption of uniform acceleration: in order to accelerate a particle of mass $\mu$ to the speed of light we would require infinite amount of energy. In the local case, due to the simplicity of the local Green function, it is possible to consider instead a source which is initially at rest and then starts accelerating: see Bondi and Gold \cite{Bondi:1955zz} and Boulware \cite{Boulware:1979qj} for the electromagnetic case, and Ren and Weinberg \cite{Ren:1993bs} for the scalar case. Boosting such a source to a finite speed, and then taking the ultrarelativistic limit, one recovers the unphysical discontinuities on the past acceleration horizon that are otherwise absent.

Unfortunately, due to the complicated analytical form of the non-local Green function, such a construction is not feasible in this case. However, based on the above discussion we may argue that if the source never reaches the future light cone (or has never emanated from the past light cone) then there would be no such singular behavior. Alternatively, one may place the $\kappa$-sources on the past horizons as a regularization prescription.

These considerations confirm our hypothesis that the unphysical assumption of uniform acceleration leads to the pathological behavior on the past and future horizons, and any physically well-behaved source should be devoid of such artefacts. Non-local theories, such as the $\mathrm{GF_2}$ theory studied in the present work, appear to be more sensitive to the physicality of sources.

\subsubsection{``Principal values'' across acceleration horizons}
\label{sec:principal-values}

While the field is logarithmically divergent on the horizon, it is possible to show that the \emph{difference} of the field across both the past acceleration horizon ($u=0$) as well as the future acceleration horizon ($v=0$) is finite. Since this difference is taken between two diverging expressions we shall refer to it as a ``principal value.'' This principal value is known from the local case, $\ell=0$. In the local theory the field is manifestly finite on all horizons, and hence the principal value becomes a mere discontinuity. Moreover, this discontinuity only appears across the past horizon, and not on the future horizon.

In this subsection we will determine the principal values across the acceleration horizons analytically (for $\rho=0$). Since $\phi$ depends only on the coordinate ${\zeta=\sqrt{|uv|}}$, the near-horizon expressions for the functions in the integrand of Eq.~\eqref{eq:nonlocretsol} can be found by inserting ${\zeta=pq}$ (with ${p>0}$ and ${q>0}$) and expanding around ${q=0}$,
\begin{align}
C^+_{\textrm{W}}(\zeta,\bar{\zeta}) &\approx \frac{1}{2} \pi \big\{4 Y_0(\alpha\bar{\zeta}) \left[\log \left(p q \bar{\zeta}/2\right)+\gamma \right] \nonumber \\
&\hspace{12pt} +\pi (\sigma_u+\sigma_v) J_0(\alpha\bar{\zeta})\big\} \, , \\
C^-_{\textrm{W}}(\zeta,\bar{\zeta}) &\approx -4 K_0(\alpha\bar{\zeta}) \left[\log \left(p q \bar{\zeta}/2\right)+\gamma \right] \, , \\
S_{\textrm{W}}(\zeta,\bar{\zeta}) &\approx \pi (\sigma_u-\sigma_v) K_0(\alpha\bar{\zeta}) \, .
\end{align}
It turns out that the difference of these integrals between either side of the horizon is independent of the position on the horizon $p$ as well as the near-distance coordinate $q$. As a consequence, the the jumps across ${u=0}$ and ${v=0}$ reduce to the finite expressions
\begin{align}
    \Delta\phi^{u=0}_0 &= +\frac{\mu\alpha}{16 \pi }\int\limits_0^\infty \dd\bar{\zeta}\,\bar{\zeta} \big[\Ei\left(-\ell^{2N} \bar{\zeta}^{2N}\right)J_0(\alpha\bar{\zeta}) \nonumber \\
    &\hspace{40pt} -4 K_0(\alpha\bar{\zeta})\big]\;,
    \\
    \Delta\phi^{v=0}_0 &= -\frac{\mu\alpha}{16 \pi }\int\limits_0^\infty \dd\bar{\zeta}\,\bar{\zeta} \big[\Ei\left(-\ell^{2N} \bar{\zeta}^{2N}\right)J_0(\alpha\bar{\zeta}) \nonumber \\
    &\hspace{40pt} +4 K_0(\alpha\bar{\zeta})\big] \, .
\end{align}
For $N=2$ one finds the analytic expressions
\begin{align}
    \Delta\phi^{u=0}_0 &=-\frac{\mu}{2\pi\alpha}\left[1-\frac12 Q(\hat{\alpha})\right] \, , \\
    \Delta\phi^{v=0}_0 &= -\frac{\mu}{4\pi\alpha}Q(\hat{\alpha}) \, , \\
    Q(\hat{\alpha}) & \equiv\, _0F_2\left(;\frac{1}{2},\frac{1}{2};\frac{\hat{\alpha}^4}{256}\right) \nonumber \\
    &\hspace{20pt} -\frac{\sqrt{\pi}\hat{\alpha}^2}{8} \, _0F_2\left(;1,\frac{3}{2};\frac{\hat{\alpha}^4}{256}\right) \; ,
\end{align}
where $Q(\hat{\alpha})$ captures the influence of non-locality. Let us emphasize that the logarithmic divergence encountered on the horizon in $\mathrm{GF_2}$ theory precisely cancels out of this symmetric limit from both sides of the horizons. In the cases of small and large values for the dimensionless acceleration parameter $\hat{\alpha}$ one finds
\begin{equation}
    \lim_{\hat{\alpha}\to 0}Q=1\;, \qquad     \lim_{\hat{\alpha}\to \infty}Q=0\;,
\end{equation}
The latter equation shows that in the limiting case of vanishing non-locality, $\ell\rightarrow 0$ (which implies $\hat{\alpha}\rightarrow\infty$), the principal value across the future horizon ($v=0$) vanishes.

Hence, the principal value across the \emph{future} horizon is solely related to the presence of non-locality, and the principal value across the \emph{past} horizon is modified by non-locality---in the local theory it is merely a discontinuity since there are no divergences. Let us also note that the contributions due to non-locality across these horizons are equal in magnitude but opposite in sign.

It is conceivable that these non-trivial principal values remain present in the non-local retarded field even after the $\kappa$-subtraction presented in the previous subsection. This is because the logarithmically divergent term, as per Eq.~\eqref{eq:c1-constant}, does not depend on the Rindler wedge and hence cancels out of the symmetric principal value prescription presented in this subsection. The constant term, however, as per Eq.~\eqref{eq:c0-constant}, differs across the Rindler wedges, giving rise to the non-trivial principal value.

\begin{figure}
    \centering
    \includegraphics[width=0.45\textwidth]{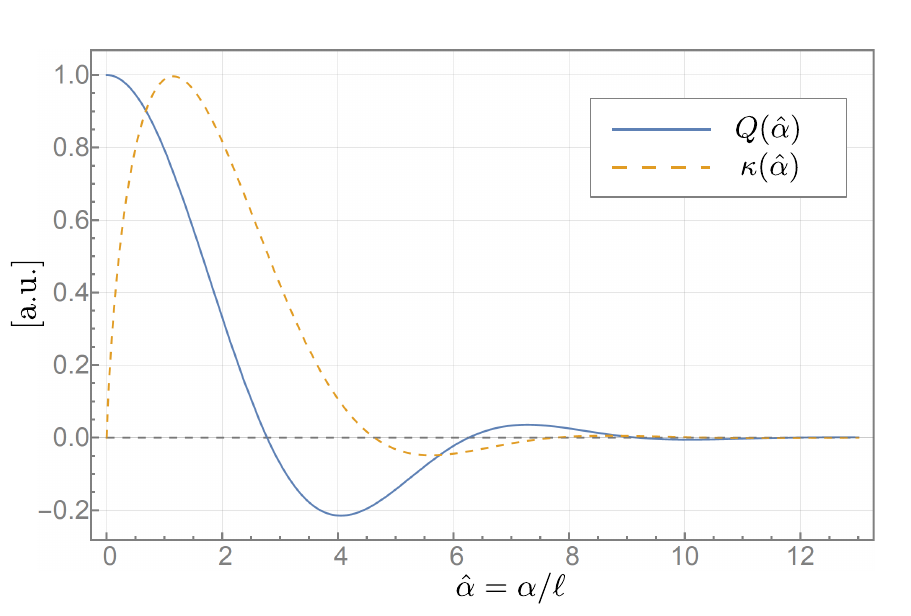}
    \caption{A plot of the dimensionless function $Q(\hat{\alpha})$ and $\kappa(\hat{\alpha})$ in arbitrary units. They both undergo non-periodic oscillations, and their zeroes do not coincide.}
    \label{fig:q-kappa}
\end{figure}

The function $Q$ exhibits damped oscillatory behavior with an infinite number of non-periodic zeroes, the first few taking place at $\hat{\alpha} = \{2.77, 6.26, 9.18, 11.81\}$. For these values the principal value vanishes across ${v=0}$. On the other hand, the quantity $\kappa$ viewed as a function of $\hat{\alpha}$ also undergoes damped non-periodic oscillations, with the first zeroes at $\hat{\alpha}=\{0, 4.63, 7.77, 10.52\}$. For those distinct values there are no divergences on the horizons, but the solution is discontinuous due to the non-vanishing principal value. A graph of the functions $Q(\hat{\alpha})$ and $\kappa(\hat{\alpha})$ can be seen in Fig.~\ref{fig:q-kappa}. Their zeroes do not coincide, which means that for select values of dimensionless acceleration $\hat{\alpha} = \alpha/\ell$ one may have either no principal value \emph{or} a finite field at the acceleration horizon.

\subsection{A non-local Born-type solution}

Before concluding, let us briefly comment on a possible extension of the retarded solution discussed so far. Namely, we would like to construct a non-local generalization of the Born solution \cite{Born:1909} and comment on its features in relation to the previously discussed logarithmic divergences and principal values.

Formally, the Born solution may be regarded as the field resulting of the retarded response of a uniformly accelerated particle in the right Rindler wedge superposed with the \emph{advanced} field of a uniformly accelerated particle in the left Rindler wedge \cite{Bicak:2002yk,Bicak:2005yt}. Instead of re-deriving Eqs.~\eqref{eq:main-calc}, \eqref{eq:ipim}, and \eqref{eq:phiIpIm} for that particular case, let us observe that we can transform the retarded field in the right Rindler wedge into the advanced field in the left Rindler wedge by mapping the null coordinates $u \rightarrow -u$ and $v \rightarrow -v$, which amounts to identifying $M_\textrm{R} \rightarrow M_\textrm{L}$ as well as $M_\textrm{F} \rightarrow M_\textrm{P}$. Let us call the retarded solution $\phi^\text{R}$ and the advanced solution $\phi^\text{A}$. The Born solution is
\begin{align}
\phi^\text{B} \equiv \phi^\text{R} + \phi^\text{A} \, .
\end{align}
The Born field in, say, the right Rindler wedge is then the superposition of the retarded field in the right Rindler wedge and the advanced field of the left Rindler wedge, and similar for all other wedges. For this reason the dependence on the factors $\sigma_u$ and $\sigma_v$, as encountered in Eqs.~\eqref{eq:main-calc}, \eqref{eq:ipim}, and \eqref{eq:phiIpIm}, drops out entirely. This immediately implies that the principal values across the horizons vanish identically for the Born solution.

The logarithmic divergences on the horizons, however, as can be seen from Eqs.~\eqref{eq:div-1}--\eqref{eq:c1-constant}, do not depend on the Rindler wedge, and hence are still present in the Born solution and need to be removed via a suitable $\kappa$-subtraction.


\section{Conclusions}\label{sc:con}

In this paper we constructed the retarded field of a uniformly accelerated point particle in a non-local scalar field theory: we employed the Sokhotski--Plemelj theorem to construct a non-local causal Green function in momentum space and found an integral representation for the resulting field. We then proved that \emph{the presence of non-locality regularizes the field at the location of the source}, while---for large timelike and spacelike distances away from the hyperbolically accelerated source---approaching the expression for the retarded field found in the local theory, in accordance with DeWitt's notion of asymptotic causality encountered in non-local theories.

On the acceleration horizons of the source, however, the retarded field is mildly logarithmically divergent due to the presence of non-locality. Using a pair of test sources on a null cone we proved analytically that such sources indeed give rise to logarithmic divergences in this particular non-local theory. We believe that this divergence is similar to those artefacts encountered in local theories, arising due to the unphysical assumption of uniform acceleration. Our considerations prove that if the source is never to become asymptotically null (either in its past or future) then there are no such divergences present, consistent with the regular field of null sources in other non-local theories \cite{Kilicarslan:2019njc,Boos:2020ccj,Dengiz:2020xbu,Boos:2020twu}. Moreover, we devise a prescription that involves test sources placed in the asymptotic past of the acceleration horizon which is capable of removing these spurious divergences. It remains to be seen if and how these additional sources are related to modified boundary conditions that one may encounter in non-local field theories. We shall leave this question for future research.

Last, we found that the difference of the retarded field across acceleration horizons is finite, even without a regularization procedure, and we demonstrated that for a non-local generalization of the Born solution these principal values vanish identically. If combined with the regularization procedure of sources in the asymptotic past one then arrives at a solution that is completely regular on the horizons.

It is a natural question to ask how the radiation of a retarded non-local source behaves, but since energy momentum tensors of non-local fields are notoriously hard to compute, see e.g.~Ref.~\cite{Boos:2019vcz} for a concrete example of $\mathrm{GF_1}$ theory, this point deserves further study. Another avenue would be the study of non-local electrodynamics, where recently ultrarelativistic objects have been studied by one of the authors \cite{Boos:2020twu}. Then, it would also be highly interesting to study implications for the presence of radiation  vis-\`a-vis the equivalence principle in Lorentz-invariant non-local theories.

Let us emphasize that the results derived in this paper present only one step towards improving our understanding of the spacetime structure of non-locality. Due to the intrinsic Lorentz invariance that lies at the very heart of this class of non-local field theories, modifications of the Green function can only be a function of the dimensionless 4-distance,
\begin{align}
\Delta\mathcal{G}(t',\ts{x}';t,\ts{x}) = \Delta\mathcal{G}\left(\frac{-(t'-t)^2+(\ts{x}'-\ts{x})^2}{\ell^2}\right) \, .
\end{align}
Naively speaking, Lorentz-invariant non-local field theories cannot seem to tell whether two points in spacetime are coincident or null-separated. Whether this presents a bug or a feature of this class of non-local theories remains to be seen.

\section*{Acknowledgements}
We would like to thank Valeri Frolov (Edmonton) for helpful comments on a previous draft of this paper. I.K. was supported by Netherlands Organization for Scientific Research (NWO) grant no. 680-91-119. J.B.\ acknowledges support by the National Science Foundation under grant PHY-181957, and is grateful for a Vanier Canada Graduate Scholarship administered by the Natural Sciences and Engineering Research Council of Canada as well as for the Golden Bell Jar Graduate Scholarship in Physics by the University of Alberta during the earlier stages of this work.

\appendix

\section{Real-space expression for the non-local modification of the scalar Green function}
\label{app:deltaG}

The free scalar retarded Green function $\mathcal{G}^\text{R}(X',X)$, due to the translational isometry of Minkowski space, depends only on the difference of its arguments, $\mathcal{G}^\text{R}(X',X) = \mathcal{G}^\text{R}(X'-X)$. Moreover, writing $X{}^\mu=(t,\ts{x})$, one can further decompose the argument structure as $\mathcal{G}^\text{R}(X',X) = \mathcal{G}^\text{R}(t'-t; \ts{x}'-\ts{x})$. A Green function in $\mathrm{GF_2}$ theory is a solution of
\begin{align}
\begin{split}
\Box e^{-\ell^4\Box^2} \mathcal{G}^\text{R}(t'-t,\ts{x}'-\ts{x}) &= -\delta(t'-t) \\ 
&\hspace{12pt} \times \delta{}^{(3)}(\ts{x'}-\ts{x}) \, ,
\end{split}
\end{align}
and clearly it is sensitive to the existence of non-locality $\ell>0$. We may decompose it as
\begin{align}
\begin{split}
\mathcal{G}^\text{R}(t'-t,\ts{x}'-\ts{x}) &= G^\text{R}(t'-t,\ts{x}'-\ts{x}) \\
&\hspace{12pt} + \Delta\mathcal{G}(t'-t,\ts{x}'-\ts{x}) \, ,
\end{split}
\end{align}
where $\Delta\mathcal{G}(t'-t,\ts{x}'-\ts{x})$ is a non-local modification term and $G^\text{R}(t'-t,\ts{x}'-\ts{x})$ is the local retarded Green function that solves
\begin{align}
\Box G^\text{R}(t'-t,\ts{x}'-\ts{x}) = -\delta(t'-t)\delta{}^{(3)}(\ts{x'}-\ts{x}) \, , 
\end{align}
subject to the retarded constraint $G^\text{R}(t'-t,\ts{x}'-\ts{x}) = 0$ if $t' < t$. From now on we shall denote $t'-t$ simply as $t$ and $\ts{x}'-\ts{x}$ as $\ts{x}$. For the local piece one may calculate
\begin{align}
G^\text{R}(t,\ts{x}) = \frac{1}{2\pi} \delta{}^{(2)}(-t^2+\ts{x}^2) \theta(t) \, ,
\end{align}
which, by construction, is only non-vanishing on the future light cone. \emph{Inside} the future light cone, as well as anywhere outside of it, it vanishes identically. The non-local part can be calculated as follows:
\begin{align}
\Delta\mathcal{G}(t,\ts{x}) &= \int\limits_{-\infty}^\infty \frac{\dd\omega}{2\pi} \int\limits_{\mathbb{R}^3} \frac{\dd^3 k}{(2\pi)^3} e^{+i\omega t - i \ts{k}\cdot\ts{x}} \frac{1-e^{-\ell^4(\omega^2-\ts{k}^2)^2}}{\omega^2-\ts{k}^2} \nonumber \\
&= \frac{\ell^2}{2\pi^{7/2} x} \int\limits_0^\infty \dd\omega\cos\omega t \int\limits_0^\infty k \dd k \sin kx \nonumber \\
&\hspace{15pt} \times \int\limits_0^\infty e^{-y^2/4} \int\limits_0^y \dd z \sin\left[\ell^2(\omega^2-k^2)z\right] \\
&= \frac{\ell^2}{2\pi^{7/2} x} \int\limits_0^\infty \dd y e^{-y^2/4} \int\limits_0^y \dd z \nonumber \\
&\hspace{20pt} \times \left[ I_1(t,z) I_2(x,z) - I_3(t,z)I_4(x,z) \right] \, , \nonumber
\end{align}
where we defined $k\equiv |\ts{k}|$ as well as $x \equiv |\ts{x}|$, and $I_1$, $I_2$, $I_3$, and $I_4$ denote the following regularized integrals:
\begin{align}
I_1(t,z) &= \lim\limits_{\alpha\rightarrow 0} \int\limits_0^\infty \dd\omega e^{-\alpha\omega} \cos\omega t \sin\omega^2 \ell^2 z \\
&= \sqrt{\frac{\pi}{8z\ell^2}}\left[ \cos\left(\frac{t^2}{4z\ell^2}\right) - \sin\left(\frac{t^2}{4z\ell^2}\right) \right] \, , \nonumber \\
I_3(t,z) &= \lim\limits_{\alpha\rightarrow 0} \int\limits_0^\infty \dd\omega e^{-\alpha\omega} \cos\omega t \cos\omega^2 \ell^2 z \\
&= \sqrt{\frac{\pi}{8z\ell^2}}\left[ \cos\left(\frac{t^2}{4z\ell^2}\right) + \sin\left(\frac{t^2}{4z\ell^2}\right) \right] \, , \nonumber \\
I_2(x,z) &= \lim\limits_{\alpha\rightarrow 0} \int\limits_0^\infty k \dd k e^{-\alpha k} \sin kx \cos k^2 \ell^2 z \\
&= \sqrt{\frac{\pi}{8z^3\ell^6}}x\left[ \sin\left(\frac{x^2}{4z\ell^2}\right) - \cos\left(\frac{x^2}{4z\ell^2}\right) \right] \, , \nonumber \\
I_4(x,z) &= \lim\limits_{\alpha\rightarrow 0} \int\limits_0^\infty k \dd k e^{-\alpha k} \sin kx \sin k^2 \ell^2 z \\
&= \sqrt{\frac{\pi}{8z^3\ell^6}}x\left[ \sin\left(\frac{x^2}{4z\ell^2}\right) + \cos\left(\frac{x^2}{4z\ell^2}\right) \right] \, . \nonumber
\end{align}
Then one can further regulate ($s^2 \equiv -t^2+x^2$)
\begin{align}
\Delta\mathcal{G}(t,\ts{x}) &= -\frac{1}{16 \pi^{5/2}\ell^2} \int\limits_0^\infty \dd y e^{-y^2/4} \int\limits_0^y \frac{\dd z}{z^2} \cos\left( \frac{s^2}{4z\ell^2} \right) \nonumber \\
&= -\frac{1}{16 \pi^{5/2}\ell^2} \int\limits_0^\infty \dd y e^{-y^2/4} \nonumber \\
&\hspace{20pt} \times \lim\limits_{\alpha\rightarrow 0} \int\limits_{1/y}^\infty \dd z e^{-\alpha z} \cos\left( \frac{s^2}{4\ell^2} z \right) \\
&= \frac{1}{4\pi^{5/2}s^2} \int\limits_0^\infty \dd y e^{-y^2/4} \sin\left( \frac{s^2}{4y\ell^2} \right) \nonumber \\
&= \frac{|s^2|}{1024\pi^2\ell^4} G{}^{20}_{03} \left( \left. \begin{matrix} \\ -\tfrac12, -\tfrac12, -1 \end{matrix} \right| \frac{s^4}{256\ell^4} \right) \, , \nonumber
\end{align}
where $G{}^{20}_{03}$ denotes a Meijer G-function \cite{Olver:2010}. It is clear that this function is invariant under $s^2\rightarrow-s^2$, meaning that it does not distinguish between timelike and spacelike distances, consistent with the putative acausality typically encountered in non-local theories.

\begin{figure}[!tb]
    \centering
    \includegraphics[width=0.45\textwidth]{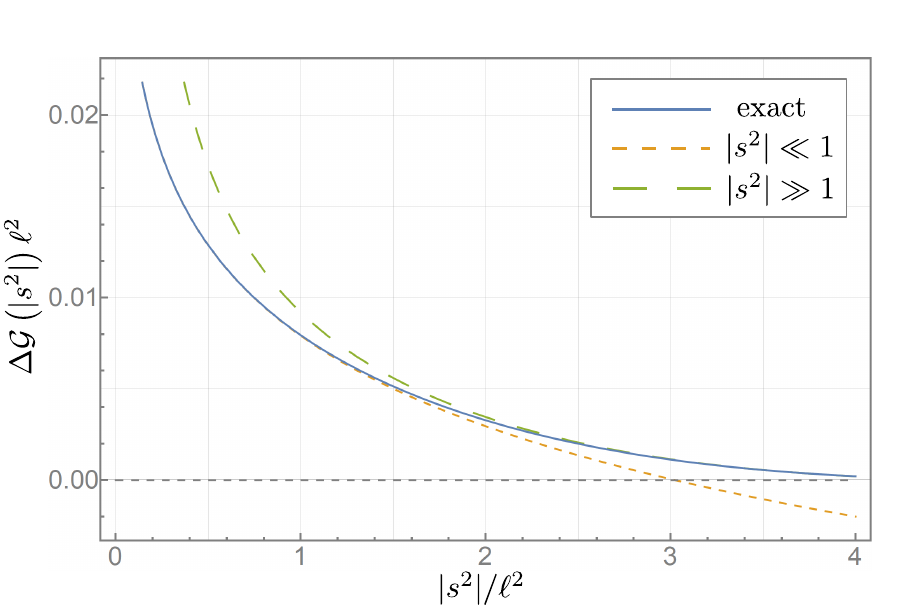}
    \caption{The dimensionless non-local modification $\Delta{G}(|s^2|) \times \ell^2$ plotted as a function of dimensionless 4-distance $|s^2|/\ell^2$, together with its null expansion ($|s^2|\ll 1$) as well as large-distance expansion $|s^2|\gg 1$.}
    \label{fig:deltaG}
\end{figure}

For small and large arguments $s^2$ one finds the following asymptotic behavior:
\begin{align}
\Delta\mathcal{G}(|s^2| \ll 1) &= \frac{1}{32\pi^{5/2}\ell^2} \left[ 2 - 3\gamma -\log\left( \frac{s^4}{64\ell^4} \right) \right] \, , \nonumber \\
\Delta\mathcal{G}(|s^2| \gg 1) &= \frac{1}{2\sqrt{3}\pi^2 |s^2|} \sin\left( \frac{3\sqrt{3}s^{4/3}}{8\cdot2^{2/3}\ell^{4/3}} \right) \label{eq:app:deltaG-asymptotics} \\
&\hspace{20pt} \times \exp\left(-\frac{3 s^{4/3}}{8\cdot2^{2/3}\ell^{4/3}}\right) \, . \nonumber
\end{align}
The non-local modification is logarithmically divergent on the light cone and decreases exponentially fast for large spacelike and timelike distances. We plot the function $\Delta\mathcal{G}(s^2)$ as well as its asymptotics in Fig.~\ref{fig:deltaG}. The exponential suppression happens in accordance with DeWitt's asymptotic causality criterion \cite{DeWitt:1965} which states that any causal Green function must satisfy
\begin{align}
\label{eq:dewitt}
\lim_{t'-t \rightarrow -\infty} \mathcal{G}(t'-t;\ts{x}'-\ts{x}) = 0 \, ,
\end{align}
that is, if the effect precedes the cause arbitrarily, any causal Green function must vanish. Since local causal Green functions satisfy DeWitt's criterion identically---since they are proportional to $\theta(t'-t)$---we only need to verify that the non-local modification satisfies condition \eqref{eq:dewitt}, which it does, as can be seen from Eq.~\eqref{eq:app:deltaG-asymptotics}.

\section{Proof of Eq.~\eqref{eq:sokplem} using the Sokhotski--Plemelj theorem}\label{sc:SPtheorem}
The Sokhotski--Plemelj theorem may be stated as follows: \textit{For any continuous function $f(x)$ one has}
\begin{equation}
    \frac{f(x-x_0)}{x-x_0\pm i\epsilon} =  \text{p.v.}_{x} \frac{f(x-x_0)}{x-x_0} \mp i\pi \delta(x-x_0)\;.
\end{equation}
These expressions are understood under the integral sign,
\begin{equation}
    \lim_{\epsilon\to 0} \int\limits_{a}^b\! \dd x \frac{f(x-x_0)}{x-x_0\pm i\epsilon} = \fint\limits_{a}^b \! \dd x \frac{f(x-x_0)}{x-x_0} \mp i\pi f(x_0)\;.
\end{equation}
First, let us consider ${\bar{X}\in M_{\textrm{R}}\cup M_{\textrm{L}}}$. Then the function ${-f(\bar{\rho})/\bar{\bs{X}}^2}$ has no poles since ${\bar{\bs{X}}^2\neq 0}$ in that region. This means that \eqref{eq:sokplem} is satisfied trivially: the Cauchy principal value integral reduces to the standard integral, and the $\delta$-term does not contribute since the momentum is spacelike. In other words, in this domain the ${i\epsilon}$-prescription is not necessary and we may simply set ${\epsilon=0}$.

If ${\bar{X}\in M_{\bar{\textrm{F}}}\cup M_{\bar{\textrm{P}}}}$, we define ${\sigma\equiv(\sigma_{\bar{u}}-\sigma_{\bar{v}})/2}$ such that ${\sigma=1}$ in ${M_{\bar{\textrm{F}}}}$ and ${\sigma=-1}$ in ${M_{\bar{\textrm{P}}}}$. Then one can show
\begin{align}
\begin{split}
&\hspace{12pt} \frac{-f(\bar{\rho})}{-(\bar{t}-i\epsilon)^2+\bar{x}^2+\bar{y}^2+\bar{z}^2} \\
&\approx\frac{-f(\bar{\rho})}{\sigma_{\bar{u}}\sigma_{\bar{v}}\bar{\zeta}^2+\bar{\rho}^2+i\bar{\zeta}\big(\sigma_{\bar{u}}e^{\bar{\tau}}{-}\sigma_{\bar{v}}e^{-\bar{\tau}})\epsilon} \\
&\approx\frac{-f(\bar{\rho})}{\Big[\bar{\rho}-\sqrt{\bar{\zeta}^2-i\sigma\bar{\zeta}\epsilon}\Big]\Big[\bar{\rho}+\sqrt{\bar{\zeta}^2-i\sigma\bar{\zeta}\epsilon}\Big]} \\
&\approx\frac{-f(\bar{\rho})}{\big(\bar{\rho}-\bar{\zeta}+i\sigma\varepsilon\big)\big(\bar{\rho}+\bar{\zeta}-i\sigma\varepsilon\big)} \\
&=\frac{1}{2\bar{\zeta}}\Bigg[\frac{-f(\bar{\rho})}{\bar{\rho}-\bar{\zeta}+i\sigma\varepsilon}-\frac{-f(\bar{\rho})}{\bar{\rho}+\bar{\zeta}-i\sigma\varepsilon}\Bigg] \\
&=\textrm{p.v.}_{\bar{\rho}}\frac{-f(\bar{\rho})}{\bar{\rho}^2-\bar{\zeta}^2}+i\sigma\pi f(\bar{\zeta})\delta{}^{(2)}(\bar{\rho}^2-\bar{\zeta}^2) \, ,
\end{split}
\end{align}
where in several lines we have rescaled $\epsilon$ by a positive constant. Utilizing this relation in Eq.~\eqref{eq:retsolcongr} and the following steps, one readily obtains Eq.~\eqref{eq:sokplem} as written in the main body of the paper.


\bibliography{references}

\end{document}